\documentclass[useAMS,usenatbib]{mn2e}
\usepackage{graphicx}
\usepackage{epsfig}
\usepackage{natbib}  
\usepackage{epstopdf}
\newcommand\aj{{AJ}}%
\newcommand\apj{{ApJ}}%
\newcommand\apjl{{ApJ}}%
\newcommand\aap{{A\&A}}%
\newcommand\mnras{{MNRAS}}%
\def\simgt{\lower.5ex\hbox{$\; \buildrel > \over \sim \;$}}
\def\simlt{\lower.5ex\hbox{$\; \buildrel < \over \sim \;$}}

\newcommand\teff{T$_{\rm eff}$}

\newcommand\Teff{T$_{\rm eff}$}

\newcommand{\msun}{\ensuremath{\, {M}_\odot}}
\newcommand{\Msun}{\ensuremath{\, {M}_\odot}}
\newcommand{\ocen}{$\omega$~Cen}

\title[A large and  extreme SG in NGC2419]{NGC~2419: a large and extreme second generation in a currently undisturbed cluster}
\author[Di Criscienzo et al.]
{M. Di Criscienzo$^{1}$, F. D'Antona$^{1}$,A. P. Milone$^{2,3}$, P. Ventura$^{1}$,V. Caloi$^4$,  \newauthor R. Carini$^{1,5}$, D'Ercole$^6$, E. Vesperini$^7$ , G.Piotto$^2$
 \\
$^{1}$ INAF, Osservatorio Astronomico di Roma, Via Frascati 33,  00040 Monteporzio Catone (Roma), Italy.\\
$^{2}$ Department of Astronomy, University  of  Padova,  Vicolo dellOsservatorio 3, Padova, I-35122, Italy\\
$^{3}$ IAC-Instituto de Astrofìsica de Canarias, E-38200 La Laguna, Canary Islands, Spain\\
$^{4}$ INAF, IASF--Roma, via Fosso del Cavaliere 100, I-00133 Roma, Italy\\
$^{5}$ Department of Physics, University ''La Sapienza'', P.le Aldo Moro 2, 00185,  Roma\\
$^{6}$ INAF-Osservatorio Astronomico di Bologna, via Ranzani 1, 1 I-40127 Bologna, Italy\\
$^{7}$ Department of Physics, Drexel University, Philadelphia, PA 19104, USA
}

\begin{document}
\date{Accepted . Received ; in original form }

\pagerange{\pageref{firstpage}--\pageref{lastpage}} \pubyear{2011}

\maketitle

\label{firstpage}

\begin{abstract}

We analyse complementary HST and SUBARU data for the  globular cluster NGC~2419. We make a detailed analysis of the horizontal branch (HB), that appears composed by two main groups of stars: the luminous blue 
HB stars  ---that extend by evolution into the RR Lyrae and red HB region--- and a fainter, extremely blue population. 
We examine the possible models for this latter group and conclude that a plausible explanation is that they correspond to a significant ($\sim 30$\%)  extreme second generation with a strong helium enhancement (Y$\sim$0.4). We also show that the color dispersion of the red giant branch is consistent with this hypothesis, while the main sequence data are compatible with it, although the large observational error blurs the possible underlying splitting.\\
While it is common to find an even larger (50 -- 80) percentage of second generation in a globular cluster, the presence of a substantial and extreme fraction of these stars in NGC~2419 might be surprising, as the cluster is at present well inside the radius beyond which the galactic tidal field would be dominant.  If a similar  situation had been present in the first stages of the cluster life, the cluster would have retained its initial mass, and the percentage of second generation stars should have been quite small (up to $\sim$10\%). 
Such a large fraction of extreme second generation stars implies that the system must have been initially much more massive and in different dynamical conditions than today. We discuss this issue in the light of existing models of the formation of multiple populations in globular clusters.

\end{abstract}


\section{Introduction}
\label{sec:intro}

Most globular clusters (GC) so far spectroscopically examined have been shown 
to contain multiple stellar populations. Together with stars having the typical 
composition of halo stars of the same metallicity, they contain a population of 
stars whose gas has been subject to the full CNO cycle (decrease of the oxygen 
content) and to proton--capture reactions on light nuclei (e.g. formation of sodium 
from neon). Taking as a probe the Na--O anticorrelation,  \cite{carretta2009a} 
show that all the 19 galactic GCs they examine display it. As the chemical 
signatures of the anomalies are present also in turnoff (TO) stars and among 
the subgiants \citep[e.g.][]{gratton2001,briley2002, briley2004}, they can not 
be attributed to ``in situ" mixing in the stars, but must be due to some process 
of self-- enrichment occurring during the first stages of the cluster life. 

Photometric evidence for the presence of multiple populations are also 
numerous, and sometimes suggestive of star formation occurring in  
successive bursts. The photometric signatures can be attributed in part to helium 
differences (morphologies of the horizontal branches (HB), multiple main 
sequences) or sometimes even to merging of different initial cluster-like 
structures \citep{carretta1851.2010}.

Different models have been discussed in the literature for the formation of 
multiple stellar populations. A general feature is that, apart 
from some exceptions, the iron content of normal and anomalous stars does not 
differ significantly \citep[according to][the iron spread in most GCs is 
contained within $\sim$0.05dex]{carretta2009ferro}. The gas having the chemical 
signatures of second generation (SG) stars must have been at least partially nuclearly processed in stars of the 
first stellar generation (FG) and it does not include supernova ejecta. 

A major problem  is the following: 
both the interpretation of the HB morphologies in terms of helium enrichment, 
and the spectroscopic information show that, in the clusters so far examined, 
the percentage of stars of the SG is generally $\sim$50--80\% 
\citep{dantonacaloi2008, carretta2009a}. Such a large fraction of SG stars can not 
be the result of chemical evolution within a ``closed box", simply because the 
processed matter available from more massive stars is always a small percentage 
of the FG mass. Anomalous IMFs of the FG do not solve the problem, instead it 
is required that the matter forming the SG stars is collected from a much 
larger stellar ensemble than what can be inferred by extrapolating to larger 
masses the present day mass function of FG stars. Possible models imply that 
either the GC has formed in the environment of a dwarf galaxy now dispersed 
\citep[e.g.][]{bekki-norris2006}, or the SG has formed in the core of a much 
more massive FG cluster, filling its tidal volume and losing about 90\% of its 
initial mass \citep{dercole2008} in the first phases of its dynamical evolution. 
In all cases, a multiple--generation cluster is today believed to be  `small' remnant of the evolution of a stellar system initially much more massive (see also Vesperini et al. 2010 for a study of the cluster properties required for the formation of a significant SG population and the implications  for the contribution of globular clusters to the Galactic halo assembly).

Important clues to understand the GC formation might come from the 
analysis of the stellar content of GCs that evolved in isolation since their formation and did not undergo any significant loss of stars that altered the initial relative numbers of FG and SG stars. 
In this work, we examine NGC~2419, one of the most massive galactic 
GCs. The cluster core relaxation timescale is 
$\sim 10^{10}$yr, while the half mass relaxation time is much longer than the 
Hubble time \citep[][2010 edition]{harris1996}. 
It is very distant ($\sim$87.5kpc according to Di Criscienzo et al. 2011), its present truncation radius is 8.74' ($\sim$220pc) according to \citet{trager1995}, while its tidal (Jacobi) radius is $\sim$700\ pc (see Sect.7).  Had this cluster evolved in its current environment since its formation, it might have retained memory of the conditions imprinted by the formation processes. On the other hand, NGC~2419 has a very extended HB morphology, with the presence of a well populated blue hook, generally regarded as a sign of the presence of an extreme, helium rich population \citep[e.g.][]{lee2005,dantona2002}. 

We take advantage of two rich HST and SUBARU observational samples, discussed in Section 2, and of new models computed on purpose for this work (Section 3) to make a detailed analysis of the HB population (Section 4), of the giant branch (Section 5) and of the main sequence (Section 6). We find several compelling reasons to 
suggest that this cluster contains two stellar generations,  with a very helium rich SG comprising $\sim$ 30\% of the cluster present population. In Section 7 we discuss this result, and what it implies for the formation and dynamical history of this cluster. 

\section{Observational data samples}
\label{sec:obs}
%
The observations consist in three different data sets including the Hubble Space Telescope ({\it HST}) images described in Sect.~\ref{subsec:hst}, and ground-based images taken with the SUBARU telescopes and described in Sect.~\ref{subsec:ground}. The main properties of these images are listed in Table.~\ref{tabelladati}.
\subsection{The {\it HST} data set}
\label{subsec:hst}
To analyse the ACS/WFC images we followed the recipes described in
\citet{anderson2008}. We used a software that analyses all the exposures
simultaneously and generates a single list of stars.
Stars are measured independently in each image by using a spatially varying
9$\times$10 array of empirical ``library point-spread functions (PSFs)'' from \citet{anderson2006}, plus a spatially constant perturbation for each exposure that
accounts for variations in the telescope focus.

The software was designed to work well both in the crowded central regions
 of the cluster and in the external uncrowded field and is able to measure
almost all the stars that would be detected by eye.
 The photometry was calibrated into the
ACS Vega-mag system by adopting the zero points given in Sirianni et al.\ (2005)
and following the procedure by Bedin et al.\ (2005).
To exclude stars that are poorly measured we followed the
selection procedures given in Milone et al.\ (2009) and
included in the analysis only relatively isolated, unsaturated stars with good
values of the PSF-fit quality index and small rms errors in photometry
and in astrometry.

Breathing can change the focus of the telescope and introduce
 small spatial variations of the PSF, which is not compensaded for in our PSF model and can result in small spatial
variations of the photometric zero point.
To minimize the effect of
any residual PSF variation on the photometry
we used the method  adopted by Milone et al.\ (2010).
 Briefly, we first determine the fiducial main
sequence for the cluster and compute for each star its color residual from it.
Then, we correct the star's color by the difference between its color
residual and the mean of those of its best-measured neighbors.
The spatially dependent correction was lower of 0.005 mag,
and accounts for both differential reddening and inaccuracies in the PSF model, regardless of the cause. The above procedure works well also in the case of a spread out MS, or even of a double or multimodal MS as shown for example  in Milone et al.\ (2010), where this same technique was applied to the double MS in NGC~6752.
\begin{table*}
\scriptsize{
\begin{tabular}{lcccc}
\hline\hline   DATE & N$\times$EXPTIME & FILT & INSTRUMENT & PROGRAM  \\
\hline
  Sep 25 2002 		     & 2$\times$400s   & F435W  & ACS/WFC   &   9666     \\
  Sep 25 2002 		     & 2$\times$340s   & F475W  & ACS/WFC   &   9666     \\
  Sep 25 2002 		     & 2$\times$360s   & F555W  & ACS/WFC   &   9666     \\
  Sep 25 2002 		     & 2$\times$338s   & F606W  & ACS/WFC   &   9666     \\
  Sep 25 2002 		     & 2$\times$340s   & F775W  & ACS/WFC   &   9666     \\
  Sep 25 2002 		     & 2$\times$338s   & F814W  & ACS/WFC   &   9666     \\
  Sep 25 2002 		     & 3$\times$340s   & F50LP  & ACS/WFC   &   9666     \\
\hline
  Dec 2002(4nights)                   & 165$\times$30/180s &  V  & SuprimeCam/SUBARU&-\\    
  Dec 2002(4nights)                   & 16$\times$30/180s  &  I  & SuprimeCam/SUBARU&-\\      
\hline
\hline
\end{tabular}
}
\label{tabelladati}
\caption{Description of the archive data sets used in this  paper. }
\end{table*}

\subsubsection{Artificial stars}
\label{subsec:hst}

 The  artificial-star (AS)  experiments which  have been  used  in this
 paper are  also fully described in  \citet{anderson2008}.   Briefly, for
 each cluster,  we generated  a list of  $10^{5}$ stars located  on the
 entire ACS field of view, with  a density that is flat within the core
 and declines as  $r^{-1}$ outside of the core.   The programs described
 in \citet{anderson2008} allow  ASs test to be performed for one star
 at a  time  and  entirely in the software: following this procedure, 
 artificial stars never interfere with each  other.  
 The ASs have a flat luminosity
 function in F606W,  instrumental magnitudes from --5 to  $-14$\ and colors
 that place them along the horizonal branch.
\footnote{The instrumental magnitude is calculated as -2.5 log(DN), where DN is the total digital counts above the local sky for the considered stars}

 The AS routine measures the images
with  the same procedure  used for  real stars.  We  considered an
artificial star as recovered when the  input and the output fluxes differ
by less than 0.75 magnitudes and the positions by less than 0.5 pixel and applied to the AS
catalog the same criteria of selection adopted for real stars.

Since  completeness  depends  on   crowding  as  well  as  on  stellar
luminosity, we measured it applying the procedure described in Milone et al.\ (2009) that accounts for both 
the stellar  magnitude and the  distance from the cluster  center.  Briefly we
divided the  ACS field into 7  concentric annuli and,  within each of
them, we examined AS results in 9 magnitude bins, in the interval $-14
< m_{\rm F814W}< -5$.  For each of these 9 $\times$ 7 grid points we
calculated the completeness  as the ratio of recovered  to added stars
within that  range of radius and  magnitude.  This grid  allowed us to
estimate  the completeness  associated  to any  star  at any  position
within the cluster. This completeness analysis is used in Section 4.2 to derive the relative  percentage of stars on  HB.

\subsection{The ground-based data set}
\label{subsec:ground}
\citet{ripepi07}, using three different data set (SUBARU, TNG and HST) 
published a CMD of NGC~2419 which, for the first time, is both deep and covers 
a very large field of view around the cluster. In particular the large field of 
view of the Suprime-Cam (34 $\times$ 27 arcmin$^2$), covered by a mosaic of 10 
CCDs and by dithering of the telescope pointings, resulted in the survey of a 
total area of 50 $\times$ 43 arcmin$^2$ centered on NGC~2419 and include both the  TNG and HST fields. In the SUBARU data, the cluster is centered on chip \#2 and it is totally covered by the five adjacent chips; the data consist of 30s and 180s exposures in  both the 
$I$ and $V$ bands. The techniques used to perform  both the  PSF photometry and 
the absolute calibration are described in detail in \citet{dicriscienzo2011}. The samples 
cover a  total range of about 9 magnitudes, from the tip of the cluster red 
giant branch   (RGB) around $V \sim$ 17 mag, down to $V \sim$ 25.7 mag, about 
2.3 mag below the TO, and extend over an area encompassing more than 1 tidal 
radius in the North-South direction and about 2 tidal radii in East-West from 
the cluster center . We have  not included in the catalogue stars lying at r 
$<$ 50'' to avoid incompleteness effects in the most crowded central regions. 
The consequence of  this conservative choice is that the typical 
internal errors of the $V$ band photometry at the level 
%
of the HB are from 0.01 to 0.02 mag. The CMD by \cite{ripepi07} shows 
that the HB of NGC~2419 extends down to an extremely long 
blue tail ending with a  ``blue hook'' \citep{whitney1998, dcruz2000}. \\ 
In addition on the same data set, \citet{dicriscienzo2011} have 
detected 101 variable stars of which 60 new variables. According to their study NGC~2419 contains 75 RR Lyrae stars, 40 of which  are located at r$>$50''. These observations   
complete the data for the brighter part of the HB, allowing  
an analysis through population synthesis\footnote{On the contrary the SUBARU data for the ''blue hook'' are largely incomplete and will not be used.}. We found that 42 of the total number of RR Lyrae stars 
found in NGC2419 by Di Criscienzo et al. 2011 are in the ACS field described in 
the  previous section. This allows us to complete with the variables 
the ACS catalogue described in Section 2.1 for the upper part of the HB.

%
\begin{table*}
\caption{HB models and isochrones}             
\label{tab}      
\centering          
\begin{tabular}{c c c c c c c c}     
\hline\hline       
Name     & Z & [Fe/H] &  $^{14}N$+p  & $[\alpha/Fe]$ & Y -HB  & Y -Isochrones & M$_{core}/M_\odot$ \\ 
\hline            
Standard & $10^{-4}$ & --2.4  & new     &  0.2    &  0.24   &  0.24  &  0.505 \\ 
          &&&&& 0.28   & 0.28 & 0.499\\
		  &&&&&& 0.35 &\\
		  &&&&& 0.42 & 0.42 & 0.466 \\
    	  &&&&& 0.7, 0.8, 0.9 & & $\geq$0.450, $\leq$0.465 \\ 
OLDN14   & $10^{-4}$ & --2.4  & old     &  0.2    &  0.24   &     \\
$\alpha$04      & $10^{-4}$ & --2.55 & new     &  0.4    &  0.25   &     \\
D2002    & $2 \times 10^{-4}$ & --2.0  & old     &  0.0    &  0.24 & 0.24 & 0.508  \\ 
&&&&& 0.28   & 0.28 &  0.497  \\
\hline 
\hline
\end{tabular}
\label{tableTEO}
\end{table*}

\section{The models}
\label{sec:models}

We computed several sets of isochrones and HB evolutions. 
Based on the paper by \cite{shetrone2001}, we choose as 
 our base composition a mixture with [Fe/H]=--2.4 and [$\alpha$/Fe]=0.2 (standard). We also 
provide a set of models for the same iron content and [$\alpha$/Fe]=0.4 ($\alpha$04). 
More recent results by \cite{cohen2010} shift this value in the range [Fe/H] from
$\sim$--2 to $\sim -2.2$. We will see that the global interpretation 
of the cluster data is not affected by the exact choice of metallicity, 
but a slightly larger iron content is
probably more adequate to understand the whole horizontal branch data (Sect. 4.2).
\begin{figure}
\begin{center}
\includegraphics[width=7.5cm]{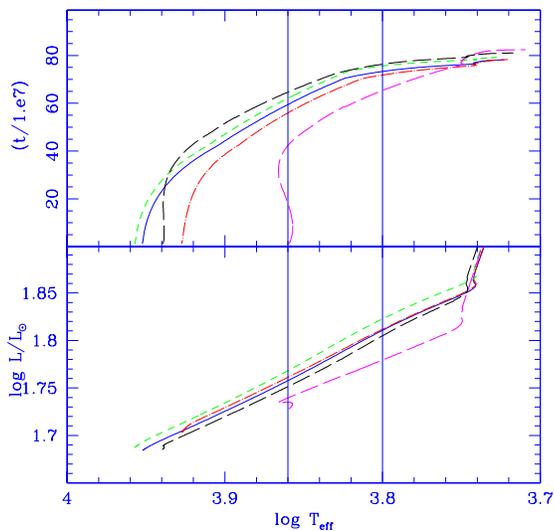}
\caption{Top panel: time evolution as function of \teff; bottom panel: HR 
diagram location of HB tracks of 0.74\Msun\ for different assumptions. Full 
line (blue): ``standard" track with [Fe/H]=-2.4 [$\alpha$/Fe]=0.2; dot--dashed 
(red): OLDN14; short dashed (green): $\alpha$04; long--dashed (magenta): 0.74\msun\ 
from D2002. The other long--dashed track on the left (black) is the 0.70\msun\ 
from D2002. } \label{f1} \end{center} \end{figure}

For T$>$10000K we used the OPAL opacities, in the version documented by 
\cite{iglesias1996}, with all the recent updates; at lower temperatures the 
opacities by \cite{ferguson2005} were adopted. Conductive opacities were taken 
from Poteckhin (2006, see the WEB page www.ioffe.rssi.ru/astro/conduct). We 
adopt the new cross section for the reaction $^{14}N$+p \citep{formicola2004}, 
but provide a set of models also for the old cross section \citep{angulo}. 
Finally, some comparisons with the data are also made with the previous models 
described in \cite{dantona2002} (hereinafter D2002) and \cite{dantonacaloi2008} computed for a metallicity 
Z=2$\times$10$^{-4}$, and no $\alpha$ enhancement. The description of the main 
inputs and of the corresponding models are given in Table \ref{tableTEO}.

We followed the main sequence and red giant evolution of low mass tracks to 
provide isochrones for the chosen chemistry and the helium core mass at flash 
for each set. The red giant evolution was followed by adopting the Reimers' (1975) mass loss law, with the Reimers parameter fixed to $\eta$=0.3. 
These core masses are also given in Table \ref{tableTEO}, and are taken as input for the 
computation of the HB models. The small helium increase in the envelope during 
the red giant evolution is also taken into account in the HB models.
Further HB models were computed, with reduced helium core mass, 
in order to simulate the effect of late flash mixing. Models of low total
mass (M$<$0.5\msun), for envelope helium
Y=0.7, 0.8 and 0.9 were added to the main HB mass sets.

Models include non instantaneous mixing of the chemicals 
species within the convection zones. We follow the diffusive approach by 
Cloutman \& Eoll (1976), solving 
for each chemical species the diffusive-like equation:
$$
{dX_i\over dt}=\big( {\partial X_i\over \partial t} \big)_{nucl}+
{\partial \over \partial m_r} \big[ (4\pi r^2\rho )^2D{\partial X_i \over \partial m_r} \big]
\eqno{(1)}
$$
where D is the diffusion coefficient, for which, given the convective velocity
$v$ and the scale of mixing $l$, a local approximation ($D\sim {1\over 3}vl$) 
is adopted. The borders of the convective regions are fixed according to the
Schwarzschild criterion. We include extra-mixing from all the formal
convective boundaries up to the beginning of the AGB phase: convective 
velocities are assumed to decay exponentially with an e-folding distance 
described by the free-parameter $\zeta$, that was set to $\zeta=0.02$,
according to the calibration provided in Ventura et al. (1998), where the
interested reader can also find a complete discussion regarding the variation 
of the convective velocities in the proximities of the convective borders.
The treatment of overshooting is particularly relevant for the HB models,
in which overshooting mimicks semi--convection \citep[e.g.][]{caloimazzitelli1990}.

The models with Z=10$^{-4}$ and Y=0.24 give an evolving mass of 0.740$\pm$0.008\msun\ to fit the location of the upper HB. Therefore we compare in Fig. \ref{f1} the evolutionary tracks for this mass and different inputs. The biggest difference is that the 
old D2002 tracks, corresponding to Z=$2 \times 10^{-4}$\ and no $\alpha$\ 
enhancement, are considerably cooler than the standard tracks. The best correspondence is found between the standard 0.74M$_{\odot}$ 
track and a track of 0.70\msun, also plotted in the figure. Doubling $[\alpha/Fe]$, but leaving the 
total metallicity unchanged, we find that the track of 0.74\msun is only slightly cooler and more luminous.
On the other hand, the $^{14}N$+p cross section does not look  to be particularly relevant for this metallicity \citep[but see
also][]{ventura2009,pietrinferni2010}. 

Synthetic models for the HB are computed according to the recipes described in \cite{dantonacaloi2008}.
We adopt the appropriate relation between the mass of the evolving giant $M_{RG}$ and the age, as function of helium content and metallicity. 
The mass on the HB is then: 
\begin{equation} 
M_{HB} =  M_{RG}(Y,Z) - \Delta M 
\end{equation} 
$\Delta M$\ is the mass lost during the RG phase. We assume that $\Delta M$\ 
has a gaussian dispersion $\sigma$\ around an average value $\Delta M_0$\  and 
that both $\Delta M_0$\  and $\sigma$\ are parameters to be determined and {\it in
principle} do not depend on Y.  Once chosen Z and Y, the  \teff\  location of an HB mass is fixed. Consequently, different ages can be adopted, provided that the mass loss is consistently adjusted.

The RR Lyrae are identified as those stars that, in the simulation,
belong to the \Teff\ interval $3.795 < \log T_{\rm eff} < 3.86$. Their
periods are computed according to the pulsation equation (1) by Di Criscienzo, 
Marconi \& Caputo (2004).

Synthetic models for the RGB and for the turnoff region are computed according to
the method described in \cite{dicriscienzo2010}.

We notice that all the evolutionary sequences have been translated from the theoretical Hertzsprung-Russell diagram (H-R) diagram to the observational CMD by using the color-Teff relation and bolometric corrections by Dotter et al. (2007). Obviously the  synthetic spectra used for these computations do not represent accurately the peculiar atmospheres of hot flashers, because these stars are expected to have strongly enhanced He and C abundances at the surface. More in detail, for the atmospheres of hot flashers, Brown et al. (2001) have shown that an atmospheric composition of 96\% He and 4\% C (or N) produces lower fluxes in the F435W filter, compared to a standard metal mixture. Therefore, for the hot flasher model, one should use more appropriate- but not yet available- bolometric corrections, that however we do not expect to alter the basic conclusions of this analysis. As for the He-rich stellar population, Girardi et al. (2007) have shown that the effect of an enhanced He content-of the order of $\Delta$Y $\sim$0.1-- 0.2 on bolometric corrections and colors is negligible at the Teff values corresponding to hot HB stars.

\begin{figure}
\begin{center}
\includegraphics[width=7.5cm]{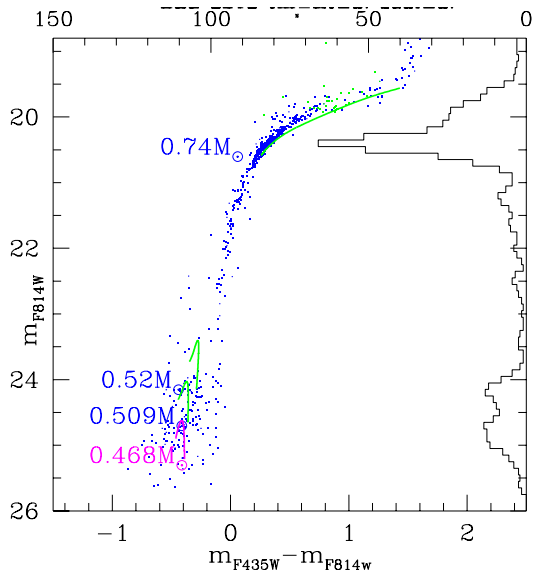}
\caption{The HR diagram of the HB in the ACS data is shown.The histogram on the right shows the star 
number distribution versus magnitude $m_{\rm F814W}$. 
Standard tracks of M=0.74\msun, M=0.52 and 0.509\msun 
for Y=0.24 and M=0.468\msun for Y=0.42 are overplotted. } \label{f3} 
\end{center} \end{figure}
\begin{figure}
\begin{center}
\includegraphics[width=7.5cm]{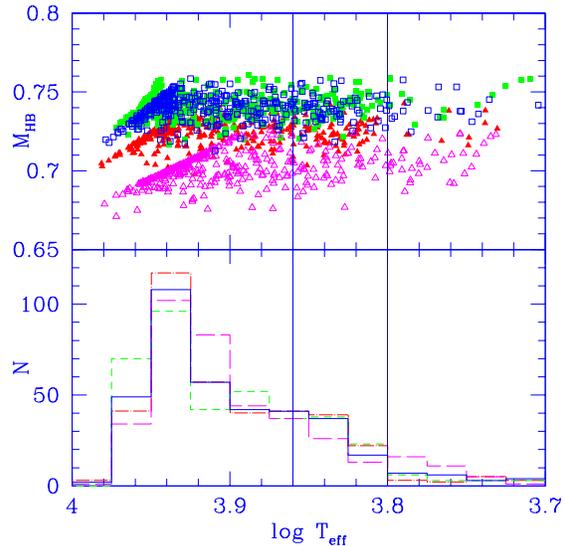}
\caption{The top panel shows the distribution of the HB masses in the different 
simulations of the upper HB. The bottom panel gives the number distribution 
versus \teff. Open squares and full line (blue): standard tracks; full squares, 
dashed line (green): OLDN14; full triangles, dot--dashed line (red): $\alpha$04; open 
triangles, long--dashed line (magenta): D2002. } \label{f2} \end{center} 
\end{figure}
\begin{figure*} \begin{center} 
\includegraphics[width=7.5cm]{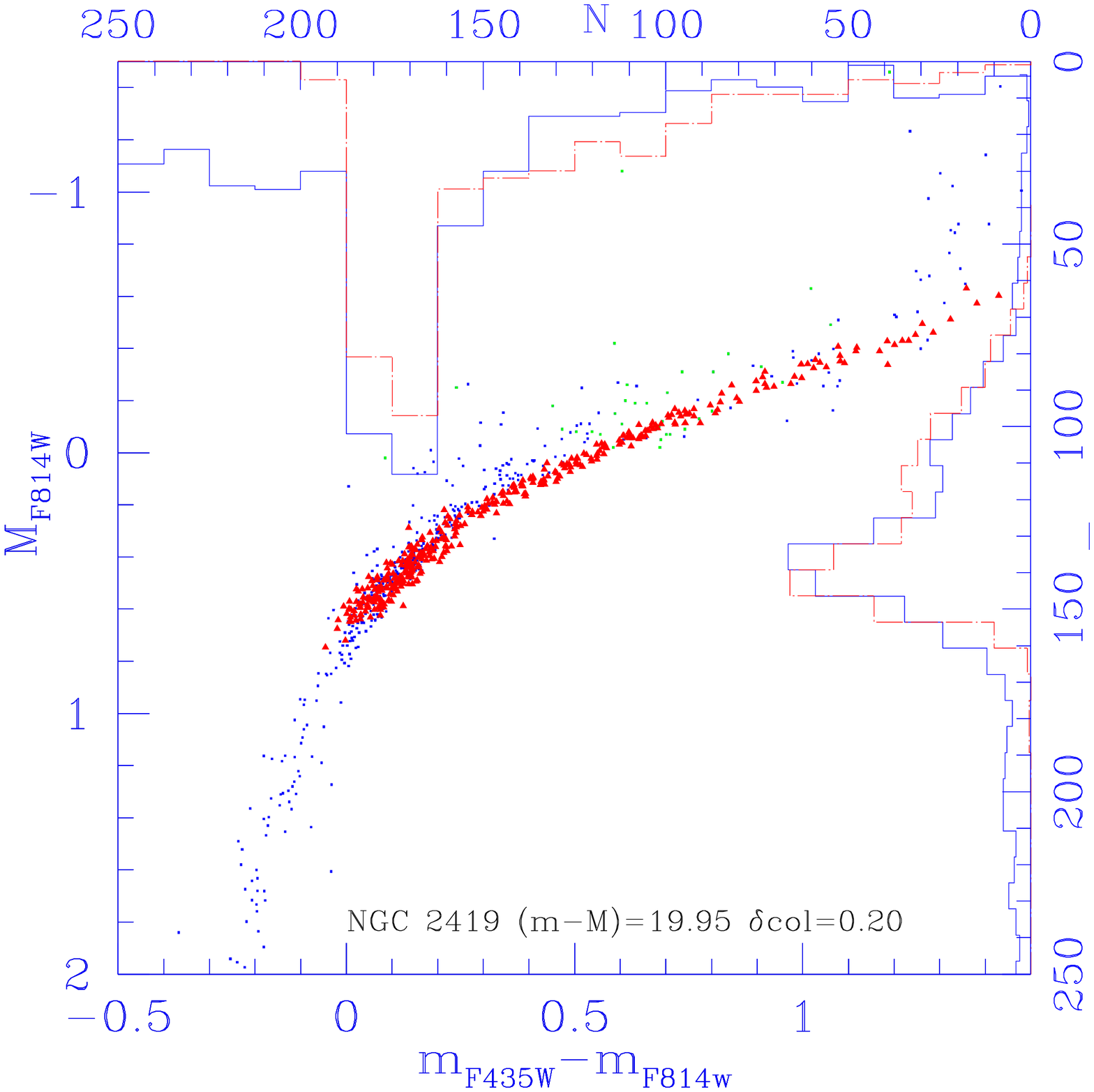}
\includegraphics[width=7.5cm]{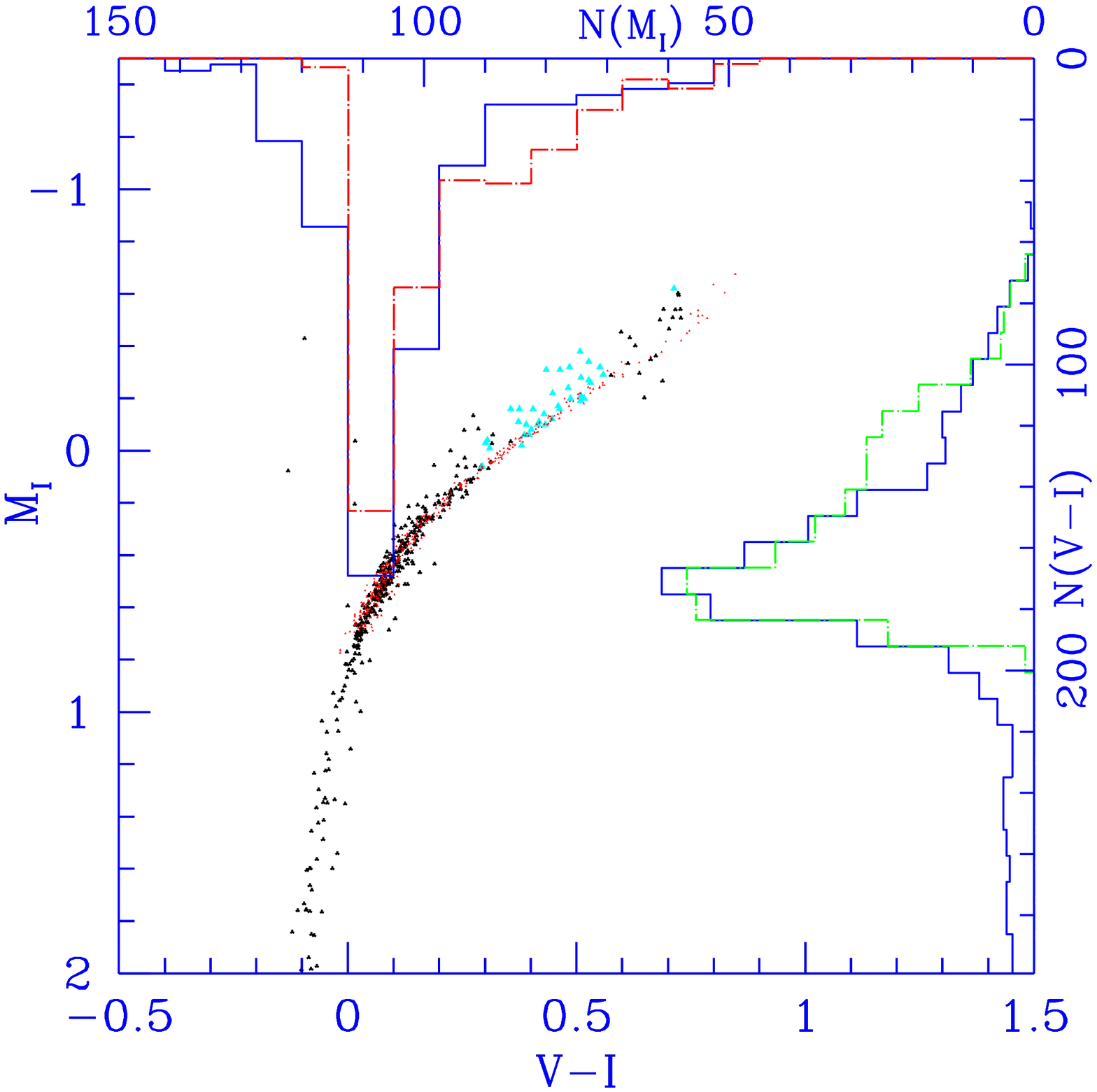} 
\caption{Simulation of the UPPER HB of NGC 2419 in the SUBARU magnitudes (left panel) and in the ACS magnitudes (right). In each figure the histograms of star counts as a function of the absolute magnitudes M$_{F814W}$\ or M$_I$\  and of the reddening corrected color m$_{F814W}$-m$_{F435W}$\ or V--I  are compared to the histograms of the simulations. A small mass spread of 0.008\msun  has been assumed. Notice that the morphology of the tracks does not allow to account for the ``blue tails", unless we postulate an asymmetric mass loss. The assumed color excess ($\delta$col) and the apparent distance modulus are also reported.}
\label{f4} \end{center} \end{figure*} 
\begin{table*}
\caption{HB counts}             
\label{tab}      
\centering          
\begin{tabular}{c c c c c c}     
\hline\hline       
 Range            & SUBARU (obs)  &  ACS (obs) & ACS(corr) & SH2008 (corr) & D'A2008 (corr)     \\ 
             & r$>$50" $\chi <$1.2& flag=1      & flag=1,c$>$0.5,r$>$50"               &      &\\
\hline            
UPPER (M$_I<$21)  & 334  & 351     & 95&   461  &  446 \\
RR LYRAE            &  40  &  42     & 12&    42 &    42 \\
MIDDLE            &  70  &  80     & 30&   130  &   53  \\
EHB               &  61  & 160     & 74&   363  &  257  \\
TOTAL (N$_{tot}$) & 504  & 633     & 211&   996  &  798  \\
\hline 
N$_{EHB}$/N$_{tot}$ &  0.12 & 0.25    &  0.35  &    0.36 &  0.34  \\
BR(EHB)/BR(TOTAL HB)&&& 0.27 -- 0.31  & 0.27 -- 0.31 &   \\
\hline
\hline
\end{tabular}
\label{tableOBS}
\\
\end{table*}
\section{The horizontal branch}
\label{sec:hb}

The HB of NGC 2419 has been analyzed in several recent studies \citep{ripepi07,sandquist2008,dalessandro2008}. We have now 
the possibility to examine further its characteristics, including in the study 
the full RR Lyrae catalogue described by \cite{dicriscienzo2011}. The HB is 
composed of two main parts: a luminous (m$_{F814W}$$<$21mag) blue section, 
with a tail extension in the RR Lyrae and in the red part  (UPPER HB), 
and the extreme HB (EHB) subluminous 
hot stars, defined here as the group at m$_{F814W}>23.5$mag, 
that probably includes both the B subdwarfs at \teff$>$20000K, and the
 ``blue hook" stars
in the standard definition related to their appearance in the UV 
\citep{whitney1998,dcruz2000,moehler2002}. 
In the middle, these two parts are connected by a tail of
intermediate luminosity stars (MIDDLE HB). The relative numbers in these three 
sections are given in Table \ref{tableOBS}. We point out that 
the UPPER HB and the EHB are the dominant components. In the following we use 
synthetic populations based on the models described above, in order to 
understand the main features of these two classes of objects.

\subsection{The luminous part of the HB: a unique population with a small spread
in mass}
\label{sec.upper}
The histogram of the number of stars as a function of m$_{F814W}$ in
Fig. \ref{f3} shows that the upper HB is extremely peaked in color and in 
magnitude. This feature has been found also in M~53 \citep{rey1998}, but not in other 
low metallicity clusters, such as M15 \citep{bingham1984}, M68 \citep{walker1994},
and strongly suggests that this portion of the HB is populated by stars spanning a narrow mass range. Figure \ref{f2} shows the 
mass distribution obtained in our simulations as follows. For any choice of the evolutionary track set, a very small mass 
spread is sufficient to describe the peaked distribution in color (or \teff). 
In fact, the RR Lyrae and red HB region, scarcely populated, are accounted for by the late evolution of the 
masses starting their HB life near the peak in the star number distribution. Notice also that the evolution at these low 
metallicities remains close to the ZAHB for all the interested masses, even in 
the Z=$2 \times 10^{-4}$ \ models of D2002. 
The simulations for this UPPER HB are done both in the SUBARU 
and in the ACS magnitudes, whose data are described in Section \ref{sec:obs}. The 
panels in Figure \ref{f4} show some results for the standard tracks and for both 
set of data. We fixed the cluster age at 12~Gyr, at which the mass evolving in 
the red giant branch is 0.813\msun. An average mass loss of 0.073\msun\ along 
the red giant phase then leads to an average mass of 0.74\msun\ in HB. A small 
spread in mass ($\sigma$=0.008\Msun) is enough to provide a good fit of the 
UPPER HB, considering small observational errors in each magnitude of 0.01mag. 
Both the SUBARU (left panel) and ACS (right panel) data are reasonably fit\footnote{We recall that we have considered 40 RR Lyrae in the SUBARU sample (only those located at r$>$50'') and 42 variables  in the ACS  field. The numbers of HB stars in the two fields, obtained adopting the same selection criteria are reported in Table 2}. 
The simulation of the SUBARU data shows a larger number of RR Lyrae  stars than found in the observations, possibly because the variable star catalogue is not yet complete. We did not investigate further this problem. In any case, we find that the period distribution of our RR~Lyr sample can be satisfactorily interpreted in terms of a ``tail distribution", in the same way as the color distribution discussed above. 

One interesting feature of these simulations is that we can not reproduce the 
extension of the HB towards  fainter magnitudes, without adding
hypotheses: any symmetric increase in the 
mass loss spread not only does not account for the long tail of stars at m$_I 
>$21, but does no longer allow a good fit of the peak. The tail ---that however 
concerns only a relatively small fraction of stars--- may be ascribed to some 
asymmetry in the mass loss, or to a relatively small enhancement in the helium 
content of these few stars (see later).  
By changing the set of tracks (see Table\ref{tableTEO}) in the simulations of the UPPER HB, the qualitative behaviour does not change. This part of the HB can still be interpreted as populated by the  evolution of a small interval of masses, around the one which provides the fits of the peak in magnitude and color.\\
Notice that a change in the cluster age does not necessarily require a change in the mass evolving on the HB, since a smaller (larger) mass loss along the RGB can be assumed for older (younger) ages.\\
The adopted metallicity (or exact elemental distribution)  do change the evolving mass: e.g. the D2002 set (Z=$2 \times 10^{-4}$)needs an average mass loss of 0.113\msun, 0.04\msun\ larger than the mass lost 
from the standard tracks (see Fig. \ref{f2})\footnote{We will see that this mass loss increase allows a better interpretation of the EHB as a group progeny of very helium rich progenitors.
}.

\subsection{The percentage of stars and the helium content in the EHB}

In this work, we will not attempt to make precise fits of the EHB population,
as the blue and near infrared bands that we are analysing are not the best
on which to accomplish such a comparison, best performed in the
ultraviolet bands \citep[e.g.][]{brown2001,lee2005,brown2010}. 
The aim of our simulations will be 
to understand, from the percentage of stars in this group, the relative birthrates
of the UPPER HB and the EHB. 
According to our completeness analysis on ACS data \footnote{We have excluded stars 
within 50'' from the center, and having a value of completness $<$50\%.} described 
in Section 2.1.1, the EHB contains about 35\% of the total HB stars.
This is consistent with the value 36\% obtained by  \cite{sandquist2008} in their
careful analysis, whose counts are reported in the fifth column of Table 
\ref{tableOBS}. 
The large fraction of stars rules out binary evolution for several  reasons: 1) typical binary fraction in luminous GCs  as NGC 2419  does not exceed $\sim$4\%  (Milone et al. 2008); 2) observationally, no binaries have yet been found among the EHB stars \citep{monibidin2006, monibidin2008}; 3) in any case, the formation of B and  O subdwarfs is a very peculiar and rare event in the binary evolution \citep[e.g.][]{han2003}; 4) the merging formation channel, that could be perhaps invoked in the dense core regions, is ruled out, as the EHB stars are distributed everywhere in the cluster within the tidal radius.\\
If the EHB is populated by stars having the same composition and age that we used for the UPPER HB, these stars have masses of  $\sim$0.51 -- 0.52\msun, whose location in ZAHB extends to the EHB region (see Figure \ref{f3}). Consequently, the mass loss of 0.073\msun, that could account for the UPPER HB, must be increased to $\sim$0.3\msun. Notice that the standard HB lowest mass stars (M$\sim$0.506\msun, being the core mass at flash 0.505\msun) do not cover the lowest luminosities of the EHB. As we will see later, we can push these masses a bit further down,  by including the location of models that have suffered a late--flash mixing\citep{brown2001,cassisi2003,brown2010}, but the problem is not fully solved with ``standard" models, and we prefer to attempt a different solution.

A dicothomy in the mass loss remains unexplained, so that we decided to test the case of similar mass loss for the EHB stars, but including a very helium rich composition for the progenitors, so that the total mass remnant after the helium flash is very small and lies at large \teff  \citep{dantona2002}. 
We assumed for the helium abundance in the EHB stars a value
Y=0.42. As we will see, the mass loss in the RG can not be kept exactly the same
as for the UPPER HB, but must be increased by $\sim$0.035 -- 0.05\msun in order to 
reach the masses (and luminosities) of the EHB. We did not try to refine this
result: a better correspondence between the mass loss in the UPPER HB and in 
the EHB can be obtained very easily by increasing the metallicity of the models (as
discussed in Sect.\ref{sec.upper}). We see that the lowest mass 
model of the Y=0.42 standard set (M=0.468\msun) is much closer to the bottom 
magnitudes of the EHB, thanks to its smaller core mass.

Obviously, the choice Y=0.42 is a bold attempt to model the higher 
helium content of the
EHB, and must not be regarded as a ``precise" value. Notice, e.g., that this value
is higher than the maximum helium contents of super--AGB stars ejecta \citep{vd2010,
siess2010}, the possible progenitors of a very helium rich second generation in GCs.

\subsection{From the number ratio to the relative birthrates of EHB and UPPER HB stars}

We will perform detailed simulations of the EHB stars in NGC 2419 with the aim to
derive a reliable ratio of birthrates of the EHB and UPPER HB, from their number 
ratio. The simulations allow us to take into account both the different evolutionary
times of the EHB stars, with respect to the UPPER HB, and the non--inclusion in
the EHB of a possible percentage of helium rich stars that evolve directly to the
white dwarf stage.

The relative birthrate of the EHB population, ${BR_{E}}\over{BR_{\rm tot}}$, 
can be written as:

\begin{equation}
{{BR_{E}}\over{BR_{\rm tot}}}={{(1+f)N_{E}/t_{E}}\over{(1+f)N_{E}/t_{E}
+ N_{M}/t_{M}+N_{U}/t_{U} }}
\end{equation}
where E, U and M indicate the extreme, upper and middle HB, f is the fraction of
stars that, due to mass loss, can not ignite the helium flash and evolves 
directly to helium white dwarfs. The $t$'s represent the typical evolutionary times 
in each of the HB groups. The relation can be rewritten putting into evidence the
timescales' ratios:

\begin{equation}
{{BR_{E}}\over{BR_{\rm tot}}}={{(1+f)N_{E}\times t_U/t_{E}}\over{(1+f)N_{E}\times t_U/t_{E}
+ N_{M}t_U/t_{M}+N_{U} }}
\label{eq2}
\end{equation}

\begin{figure}
\begin{center}
\includegraphics[width=7.5cm]{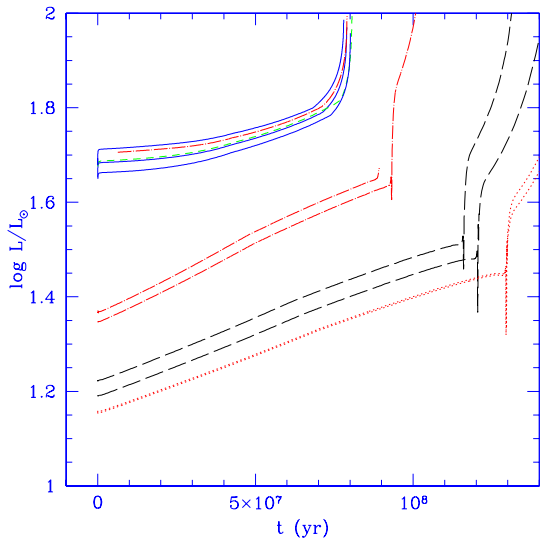}
\caption{In the upper part of the figure time evolution of standard tracks of 0.72, 0.74 and 0.77\Msun\ 
(continuous lines) plus 0.74\Msun\ OLDN14 (dot--dashed) and $\alpha$04 (short--
dashed). Dot--dashed lines in the middle are the evolution of 0.509 and 
0.52\msun\ for Y=0.24 and dashed lines below are the 0.468 and 
0.480\msun\ for Y=0.42. Dotted lines at the bottom are the evolution of
M=0.460\msun\ for Y=0.8 and Y=0.9 in the envelope.}
\label{f6}
\end{center}
\end{figure}

Figure \ref{f6} shows the luminosity versus time evolutions of typical masses
populating the HB: the 0.72, 0.74 and 0.76\msun\ standard track are taken as a reference.
Comparison with one track computed with the old $^{14}$N+p cross sections shows that
the timescales do not change significantly. In the middle, the 0.509\msun\ evolution,
for Y=0.24, shows that the evolutionary times increases by  $\sim$10\%, while the
evolution of 0.468 and 0.480\msun\ for Y=0.42, evolving at the lowest luminosity, show
an increase by $\sim$50\%, and a bit more if we considere models representing 
the result of late helium flashers (see later on). Notice that the evolutionary
times in HB depend very much on how semiconvection and/or overshooting are
treated. In our case, models are computed by fixing to $\zeta$=0.02 our parameter 
for diffusive extra--mixing (see Sect.\ref{sec:models}), and we are implicitly assuming that
the overshooting affects in a similar way both the UPPER HB and the blue hook models.

In order to have a reasonable idea about the factor {\it f}, we must make simulations,
and discuss the hypotheses hidden in these simulations. 
Although the photometry is not as good as in other clusters, we
can make the approximation that the EHB group is a mixture of HB stars progeny of 
giants that suffered an ``early" late flash after leaving the RGB, plus stars
that suffered a ``late" late flash, and have been subject to deep mixing during 
this event \citep{brown2001}. Which fraction of stars followed which path can
be extracted from comparison between the simulations
and the data. If we wish to simulate the late flash
product, the limiting total remnant mass, following the
late flash, is not the helium core mass reached at the flash occurring
on the RGB (plus a few thousands of \msun\ in the hydrogen rich envelope), 
but is smaller, as the late flash can ignite at a smaller core mass.
Following the few results in the literature \citep{dcruz1996,millerbertolami2008} 
and our own preliminary 
computations (Ventura et al. 2010, in preparation), we put a limit at the
minimum core mass for ignition of the late flash at 0.015\msun\ below the core masses
listed in Table \ref{tableTEO}. In order to simulate the late flash products, we compute models
at these very low total masses by imposing that helium at the surface is increased
to Y=0.9. In \cite{dantona2010} we proposed to fit the blue hook sequence in
the cluster \ocen\ by models with helium increased to Y=0.8, and showed that
a very interesting reproduction of the ``vertical" shape of the blue hook was 
possible. Nevertheless, the presence of carbon enhancement in the atmospheres
of blue hook stars \citep{moehler2002,moehler2007} 
links them more closely to the result of flash
mixing, so that this is the approach we follow in this work.

\subsection{Simulations of EHB}
Among the different possible hypotheses, we limit the discussion to the following
cases:
\begin{enumerate}
\item the EHB is populated by stars having the same standard chemistry 
(Y and [Fe/H]) of the UPPER HB, but subject to larger mass loss on the RGB. 
The limiting mass is a bit smaller than the helium core mass at flash, 
M$_{min}$=(0.505\msun--0.015\msun), when this occurs on the RGB.
\item the EHB is populated by stars with Y=0.42, and the minimum mass is 
fixed as in case (i), that is to M$_{min}$=(0.466--0.015)\msun;
\item stars in the EHB follow the same path as in (ii), but the fraction 
with masses smaller than 0.468\msun are considered ``late flashers", and their 
surface composition is considered to have been altered by deep mixing providing Y=0.9.
No account is taken in the models of the probable carbon enrichment to 1--3\%.
\end{enumerate}
The last case corresponds then to the typical evolution in other clusters containing
blue hook stars. 
Although our ACS sample is incomplete, we compare the
simulations to the sample as it is, as we are interested particularly in obtaining
a reliable description of the EHB sample luminosities, and in evaluating the
possible number of ``lost" helium white dwarfs, than in the completeness problems.
We will use the corrected fraction of EHB stars (Table \ref{tableOBS}) when we will try to 
understand the relative birthrates.
Therefore, a total number of 160-170 stars is assumed in the simulations.
We fix the average mass lost on the RGB ($\delta$M) and its dispersion $\sigma$.
We fix the limiting minimum mass, as discussed above, and any 
extraction ending with a smaller mass is rejected and accounted for as 
helium white dwarf. In the simulations we see that
we need to lose a bit more than the 
0.073\Msun\ taken to reproduce the UPPER HB, by 0.035--0.05\msun. We do not think 
that this is relevant, as we have seen that a slightly different choice of 
the input metallicity may allow us to require a bit larger mass loss for the 
UPPER HB, consequently reducing the discrepancy between the needed mass
loss between the UPPER HB and the EHB.

\begin{figure}
\begin{center}
\includegraphics[width=7.5cm]{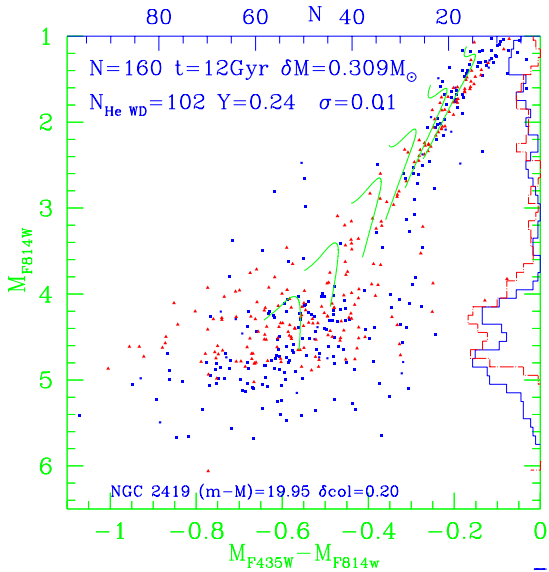}
\caption{We show the HR diagram of the EHB in the ACS data, and the histogram of the
number versus M$_{\rm F814W}$\ distribution, together with the tracks of Y=0.24 
and M=0.509, 0.52, 0.54, 0.56, 0.58 and 0.6\msun\ from bottom to top. 
The simulation (triangles and dash-dotted histogram) is described in the text.
}
\label{f24}
\end{center}
\end{figure}

Figure \ref{f24} shows one of the simulations corresponding to case (i). 
The (red) triangles and the dot-dashed histogram represent 
a simulation obtained for assuming a sample of stars with
Y=0.24, mass loss $\delta$M=0.309\msun\ on the RGB, and $\sigma$=0.01\msun. 
The limiting mass is assumed to be 0.509\Msun for the normal HB, and further the
mass extraction is extended to 0.49\msun, assuming this as the limiting mass
for the late helium flash. No mixing is included, so the hypothesis is 
quite conservative, but including models with mixing does not
improve much the fit, as these models have too high a luminosity in any case.
102 helium white dwarfs are predicted by this simulation, 
a number that we can not consider realistic. 
If we attempt to reduce the number of helium white dwarfs by decreasing
the average mass lost on the RGB, the simulated stars become too luminous with
respect to the observed EHB sample. Notice that we can not change
the distance modulus, that is obtained by fitting the UPPER HB.  
In Fig. \ref{f24}, the MIDDLE HB is also simulated, by a separate
simulation of standard composition stars with mass loss of 0.22\msun\ and
mass dispersion $\sigma$=0.05\msun. It is not possible
to reproduce in other ways these stars, neither by a simulation centered on the EHB, 
with larger mass spread, nor as an extension of the UPPER HB, as we discussed above.

\begin{figure}
\begin{center}
\includegraphics[width=7.5cm]{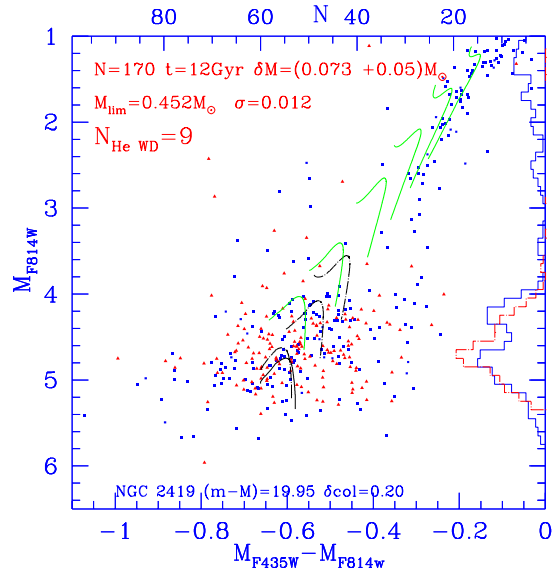}
\caption{As in Fig.\ref{f24}, the ACS data are compared to a simulation
that assumes Y=0.42 for the EHB stars. In addition to
the Y=0.24 tracks shown also in Fig \ref{f24}, three
tracks for Y=0.42 and M=0.468, 0.480 and 0.5\msun. 
The lowest track is for M=0.452\msun 
(core mass 0.45\msun) The dot-dashed histogram
represents the simulation described in the text, that assumes an average
mass loss $\delta$M=0.123\msun\ on the RGB, and $\sigma$=0.006\msun. 9 helium
white dwarfs are predicted by the simulation.
}
\label{f5}
\end{center}
\end{figure}

\begin{figure}
\begin{center}
\includegraphics[width=7.5cm]{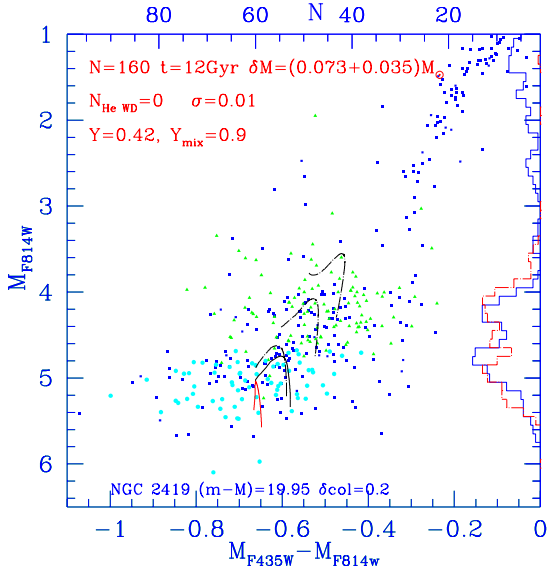}
\caption{As in Fig.\ref{f5}, but the simulation now includes 100 stars with Y=0.42 
(triangles), plus 70 stars supposed to have undergone flash mixing, having Y=0.9
(dots). In addition to the four tracks Y=0.42, the mass 0.46 for Y=0.9 is shown
(last on the bottom left). No helium white dwarfs are predicted by the simulation.
}
\label{f95}
\end{center}
\end{figure}

Figure \ref{f5} shows an example of simulations for case (ii). 
The mass spread around the average value is taken to be 
$\sigma$=0.006\msun. The total number of helium white dwarf of the simulation 
is only 9, namely $\sim$5\%.
Case (iii) is explored in Fig.\ref{f95}. The fit of the EHB sample is here very
adequate, thanks to the inclusion of the dimmer (in the F814W band) points 
corresponding to the supposed flash--mixed stars. Notice that also in this
case the numer of helium white dwarfs is predicted to be zero.

Other simulations predict a larger number of helium white dwarfs, depending on the
assumed mass loss dispersion assumed. Independently from the observational 
incompleteness, we stress that the blue hook progenitors are probably
within a maximum $\sim$20\% more than the EHB sample. 

The simulations show that {\it the EHB is better simulated including 
the presence of a large population with enhanced helium.} 
We can now go back to Equation \ref{eq2}, including the number ratios of
Table \ref{tableOBS} by \cite{sandquist2008}, f between 0 and 0.2, assuming also
$t_{E}/t_U$=1.56, $t_{M}/t_U$=1.13 from the models. If we consider the cluster ``second 
generation" to be limited to the EHB sample, the birthrate of this population is between
27 (f=0) and 31\% (f=0.2) of the total cluster population. If we include the MIDDLE
HB in the second generation, the figures increase to 40 -- 44\% of the total. The
MIDDLE HB however does not reach the very high helium contents required for the
EHB sample.

We now examine the other HR diagram features 
to understand whether this conclusion is consistent with the information 
derived from the RGB and MS.

\section{The Intrinsic Broadening of the base of  RGB} 
\begin{figure}
\begin{center}
\includegraphics[width=9.0cm]{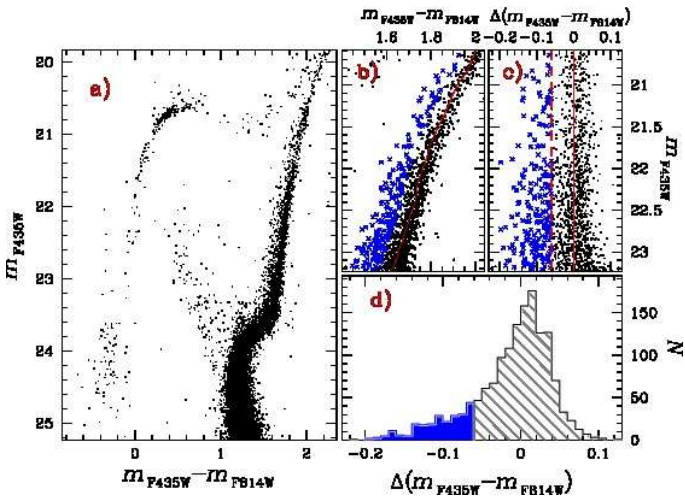} 
\caption{{\it a)}m$_{\rm F475W}$ {\it vs.} m$_{\rm F475W}$-m$_{\rm F814W}$ 
CMD  from ACS/WFC data {\it b)} and a zoom around the RGB region. 
Panels {\it c)} and {\it d)}show the analysis performed to put into
evidence the spread of the RGB in the way described in the text.}
\label{NUOVA1} 
\end{center} 
\end{figure} 
\begin{figure}
\begin{center}
\includegraphics[width=7.5cm]{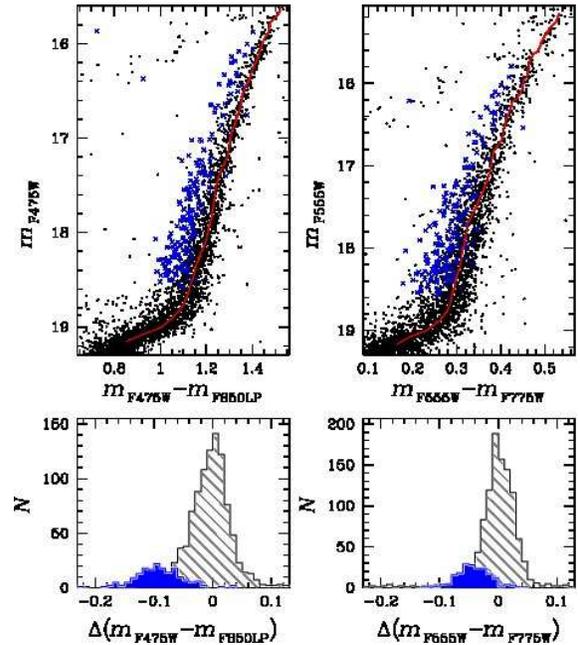}
\caption{In order to show that the RGB broadening is intrinsic we compare the CMD of Fig.~\ref{NUOVA1} with CMDs from two other data sets (see Table 1). Upper panels are the CMDs zoomed around the
RGB and the (blue) cross identify stars selected as ''blue'' in the previous Fig. 9. Lower panels show the color
distributions in these new colors. The fact that the histogram distributions of the selected blue stars sistematically have bluer colors demonstrates that the RGB broadening is intrinsic.} 
\label{NUOVA2} 
\end{center} 
\end{figure} 
\begin{figure} \begin{center} 
    \includegraphics[width=7.5cm]{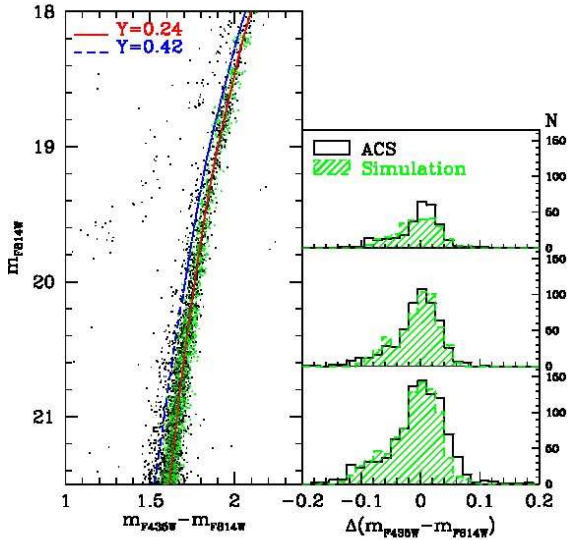} 
\caption{Zoom of the CMD region at the base of RGB from ACS data toghether with our best simulation (see text).(Right panels) Distribution of  the difference 
between the color of each star and the averaged Ridge Line calculated for this 
part of RGB calculated in three different interval of magnitude (solid 
histogram) which highlights the intrinsic broadening of the data well 
reproduced by our simulation (dashed-shaded hystogram) made up of 30 \% off stars with Y=0.42 and the rest with primordial helium abundance.
} 
\label{RGB1} 
\end{center} \end{figure} 

\begin{figure}
\begin{center}
\includegraphics[width=7.5cm]{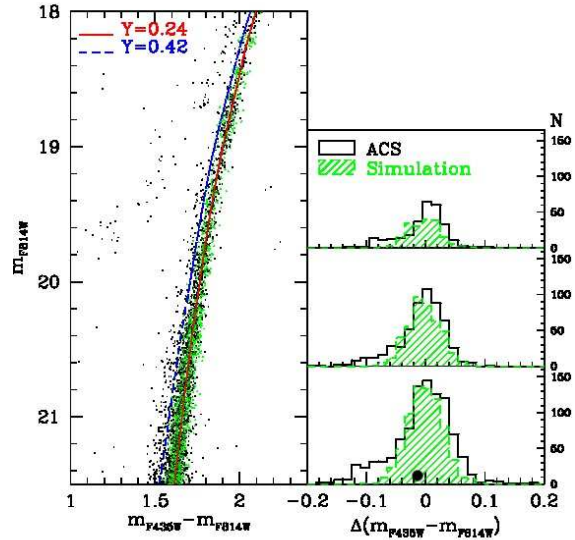}
\caption{The same as Fig. \ref{RGB1}, but in this case the simulation includes only stars with  primordial helium content (Y=0.24).} 
\label{100p} 
\end{center} 
\end{figure} 
\begin{figure}
\begin{center}
\includegraphics[width=7.5cm]{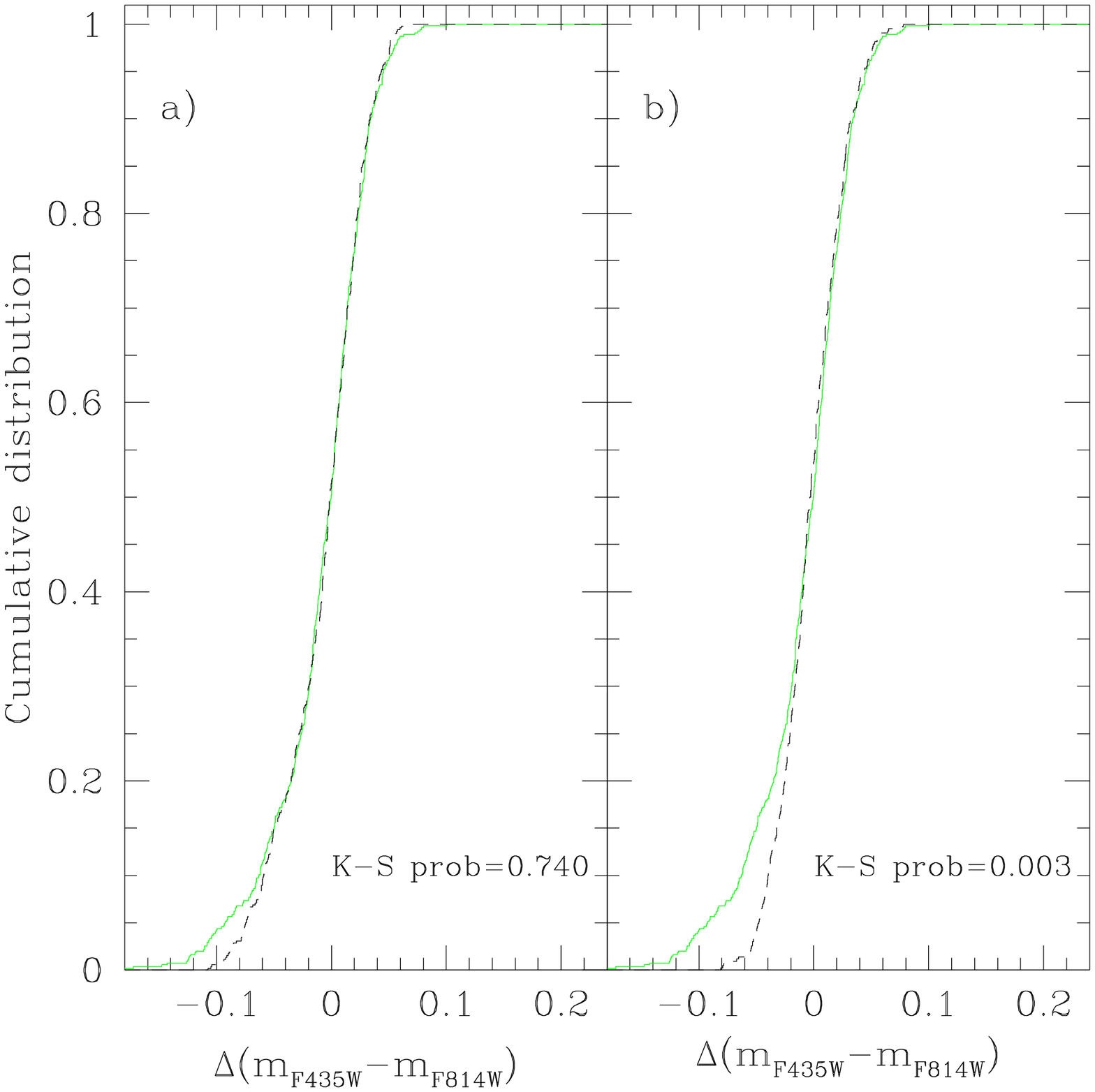}
\caption{Cumulative distributions of  $\Delta$ (${\it m}_{\rm F435W}-{\it m}_{\rm  F814W}$) from observed (dashed lines) and simulated (solid lines) stellar components computed from (a) two populations with $\Delta$Y=0.18 and (b) a single population  with fixed Y=0.24. The results of the comparison between the distributions using a K-S test (K-S prob) are also labelled.} 
\label{cum} 
\end{center} 
\end{figure} 
In order to confirm the suggestion that the blue hook stars are enhanced in helium,
we analize the stars along the red giant branch, where a large He enhancement manifests itself as a RGB split, or as a spread in a well populated CMD. Theoretical predictions \citep[e.g.][]{catelan2010} show that the base of RGB of He-enhanced models is hotter than their lower -- helium counterparts, by about $\Delta$(B-I)/$\Delta$Y=--0.7. For the case
explored here ($\Delta$Y ~0.18), this spread is larger than the observational errors for both our data sets (SUBARU and HST).
The size of the predicted split/spread  decreases progressively towards the RGB tip, where the presence of AGB stars may also complicate any empirical test. Consequently, we restrict our analysis to the interval in apparent magnitude in between the base of RGB and m$_{\rm F814W}$ $\sim$ 19 mag.\\
The main challenge in identifying intrinsic spreads in the colors of
the RGB, comes from the fact that photometric errors can
 generate similar signatures.
\citet{anderson2009} have introduced an efficient approach to
distinguish intrinsic color broadening from mere photometric errors
that consists in analysing different data sets and see if all of them
share the same features.
The large amount of ACS/WFC archive images of NGC 2419 (see Table 1)give us the
precious opportunity to follow a similar procedure for this cluster.
Specifically, we made three independent CMDs by using images in six
passbands:
i) ${\it m}_{\rm F435W}$ and ${\it m}_{\rm F814W}$, ii) ${\it m}_{\rm F475W}$
and ${\it m}_{\rm F850LP}$ and iii) ${\it m}_{\rm F555W}$ and ${\it m}_{\rm F775W}$.

The ${\it m}_{\rm F435W}$ vs. ${\it m}_{\rm F435W}$-${\it m}_{\rm
  F814W}$ CMD is shown in Fig.~\ref{NUOVA1}a. The wide color baseline
 of this CMD provide an optimal resolution of any doubling or
 spread of the observed sequences due to helium variations.
A visual inspection of this CMD immediately suggests that
NGC 2419 has a broad RGB with the presence of a tail of stars
blueshifted with respect to the main RGB population.

To further investigate this suggestion in the {\it Panel b)} we show a
zoom of the CMD around the RGB region where the color spread is more
evident. The red continuous line is the RGB fiducial obtained
by following the recipes given in Milone et al.\ (2008). Briefly, we
have divided the RGB into intervals of 0.2 magnitudes in the F435W band,
calculated the average color and magnitude for each of them, and
interpolated these points by means of a spline. In {\it Panel c)} we
have subtracted from the color of each star the color of the fiducial
at the corresponding magnitude and plotted ${\it m}_{\rm F435W}$ as a
function of the obtained color difference ($\Delta {\it m}_{\rm
  F435W}-{\it m}_{\rm  F814W}$).  The histogram color distribution plotted in
the {\it Panel d)} is clearly skewed towards blue colors and we have
arbitrarily isolated RGB stars with $\Delta {\it m}_{\rm  F435W}-{\it
  m}_{\rm  F814W}<-0.06$. These stars have been represented as blue
crosses in the {\it Panels b)} and {\it c)}.

We note that if the RGB broadening is due to photometric errors alone,
a star that is bluer than the RGB fiducial in the ${\it m}_{\rm
  F435W}$  vs. ${\it m}_{\rm   F435W}-{\it m}_{\rm  F814W}$ CMD have
the same probability of being bluer or redder in other CMDs obtained
with independent data. But if the color spread is real the selected
blue stars will have bluer colors in all the CMDs.
In Fig.~\ref{NUOVA2} we compare the CMD of Fig.~\ref{NUOVA1} with CMDs
from two other data sets. Upper panels are the CMDs zoomed around the
RGB. Each (red) line is the RGB fiducial obtained with the method
described above and we have kept for each star the same color that it
had in Fig.~\ref{NUOVA1}. Lower panels show the color
distributions. The fact that the histogram distributions of the
selected blue stars sistematically have bluer colors demonstrate the
the RGB broadening is intrinsic.
We now  can consider this evidence as the consequence of 
a population in the cluster with an enhanced primordial helium. A  
quantitative comparison is done by performing simulations able to
reproduce the color distribution. Population synthesis 
follows the outline described in \cite{ventura2009}. 
A good match is obtained by assuming a sample of 70\% of stars with 
primordial helium abundance and 30\% of stars with Y=0.42, if 
we assume $\sigma$$_{obs}$=0.018mag ( see Fig. 11).
This assumptions are compatible with the observational errors of the ACS/HST data and the simulations can reproduce both the width and the shape of distribution.\\ 
On the contrary, Fig.\ref{100p} shows that a single population with fixed Y=0.24, 
is not able to reproduce the observed population, especially in the faintest magnitude interval. 
To quantify this statement in Fig. 13 we show the cumulative distributions of the color differences ($\Delta {\it m}_{\rm F435W}-{\it m}_{\rm  F814W}$); the results of the comparison between the distributions using a K-S test are also listed which show that the agreement between observations and simulations is better if we assume a double population.\\
If we could reduce the observational errors, the split of the two RGBs would
be observable for $\sigma$$_{obs}$=0.01~mag in the same colors at least at the base of RGB; 
at the distance of NGC 2419 this will be 
possible when the next generation telescopes (e.g E-ELT or JWST) will be available.\\
We note that in the case of NGC2419 that population that we recognize as the  second generation is bluer on RGB than the primordial one, contrary to what happens for other clusters which shows evidences of multiple populations, in particular M4 (Marino et al. 2008) and NGC6752 (Milone et al. 2010) where Na rich stars are cooler  then Na poor ones. We interpret this evidence as the 
consequence of the strong He enhancement found for the SG stars
which, at the low metallicities of NGC2419, prevails on the effects of light element abundances (such as Na enhancement, or O depletion).\\

\section{The Main Sequence}
\begin{figure}
\begin{center}
\includegraphics[width=7.5cm]{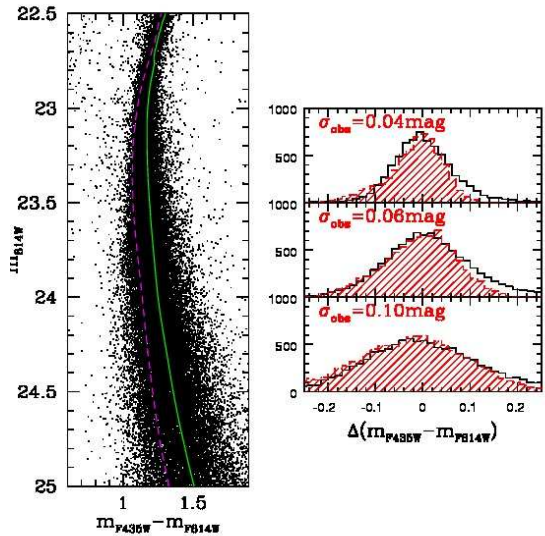}
\caption{Same as Fig. \ref{RGB1} but for three different bins in magnitude on the Main Sequence. The solid and dashed lines in the left panel refer to the isochrones calculated for Y=0.24 and Y=0.42 respectively. The observational errors reported in each box and included in the simulations   reproduce the enlargement of $\sigma$$_{obs}$ going to fainter magnitudes.}
\label{MS}
\end{center}
\end{figure}
\citet{sandquist2008} have ruled out  the hypothesis of a very high initial helium  abundance (Y $\sim$ 0.4)  for blue hook stars, mainly by noticing that the  distribution in color of the MS stars from their data resemble a Gaussian distribution and does not show any asymmetry which would arise from the presence of 30 per cent of He-rich stars. The  symmetry of the MS is confirmed by both our data sets where no bimodal main sequence  nor any noticeable asymmetry of the distribution emerge; however  our simulations show that the observational errors of our data, which increase with magnitude, may be able to  cancel any color shift.
To explain this point in Fig. \ref {MS}, we also show that  the data are perfectly matched with a synthetic population formed by  30 $\%$ of stars with $\Delta$Y=+0.18 with respect the main population.

In conclusion we notice that although Sandquist $\&$ Hess (2008) limit a maximum $\Delta$Y among the cluster stars to $<$0.05, this value is derived from the ratio R=N$_{HB}$/N$_{RG}$ where N$_{RG}$ are the giants above the HB level and not from the broadness of the MS. As well known, this ratio is not a sensitive helium indicator, if the horizontal part of the HB corresponds to the luminosity level of standard helium stars (see also Caloi $\&$ D'Antona, 2007). This is in fact the case for NGC 2419, where helium rich stars are only those in the EHB.

\section{Conclusions and discussions: why two populations?}
The fundamental result of this work is the following: in contrast with
previous results, we have built a well founded
case for the presence of two different stellar populations in NGC 2419, distant massive GC, now well isolated from the tidal influence of the Galaxy. 
The first stellar population is identified with the UPPER HB including the luminous blue stars, the RR Lyrae stars and the
few red HB stars. These have a normal He abundance and are easily
interpreted as the result of the evolution of
a small range of initial stellar masses in ZAHB. The second stellar population represents 30 per cent of the total cluster stellar population and is identified with the EHB possibly subdivided into a B--subdwarf part plus a blue hook, that we fit with models having initial helium content Y=0.42 and being
subject to a mass loss very similar to the mass loss required for the UPPER HB.
The value Y=0.42 is not mandatory, as a different, larger, 
choice of the initial [Fe/H] may allow a fit with a smaller (but in any case very high)
helium. The MIDDLE HB may be part of the helium enirched population --but with
a much smaller helium enhancement-- or may be attributed to a tail of larger mass
loss along the RGB, in standard helium stars. In the most conservative case
(only the EHB is helium rich), we conclude that 27--30\% of the cluster stars
are born from a ``second generation" very helium rich gas.\\

It is now interesting to discuss the implications of the results
presented in this paper for the initial properties of NGC2419 and its
subsequent dynamical history.\\
According to the models presented in D'Ercole et al. (2008, 2010), the
Extreme population must have formed early in the SG formation process
directly out of the pure ejecta of massive AGB stars without any
dilution with pristine gas. For example, in D'Ercole et al's (2010)
model for NGC 2808 (a massive cluster with a significant fraction of E
stars), the Extreme population observed in this cluster would have
formed in the time interval 31.7-44 Myr from the AGB ejecta of  stars  
with masses in the range 7.5-9 M$_{\odot}$.\\ 
If we assume that the Extreme population of NGC 2419 was also formed
from the ejecta of stars in this same range of masses and adopt a
Kroupa-1993 IMF (Kroupa et al. 1993), the total amount of gas lost by these
massive AGB stars is $M_{\rm ej}=0.01M_{\rm FG,i}$ , where $M_{\rm
FG,i}$ is the initial GC mass, composed exclusively by FG stars. From
the observational results presented in this paper, we can write
$M_{\rm SG,E}=0.3M_{\rm GC}$, where $M_{\rm SG,E}$ is the mass of the
Extreme population, and $M_{\rm GC}$ is the present GC mass. Assuming
that all the AGB ejecta is exhausted by the star formation, and that
no substantial loss of SG stars is suffered by the cluster during its
successive dinamical evolution, we obtain $M_{\rm SG}=0.01M_{\rm
FG,i}=0.3M_{\rm GC}$ and, in the end, $M_{\rm FG,i}=30M_{\rm GC}$. If
a Kroupa-2001 IMF (Kroupa 2001) is adopted, the larger fraction of the
total cluster mass contained in stars within the range 7.5-9
M$_{\odot}$ leads to $M_{\rm FG,i} \simeq 15 M_{\rm GC}$. 
These computations show that for any choice of  the IMF, NGC2419 lost a larger amount of its original mass than ''normal'' Galactic GCs, even up to 97 per cent.\\
The above conclusions are not surprising as a few more massive
clusters found to host E stars also have a spread of heavy elements
and must have been initially more massive by a similar large factor
(see e.g. Renzini 2008). However the unusual relative number of
Extreme and Intermediate stars in NGC 2419 is likely to be the result
of a peculiar SG formation history. Specifically, the lack of a
significant fraction of an Intermediate SG population suggests that
the SG formation episode was interrupted earlier before the ejecta of
AGB stars with $M<7$ M$_{\odot}$ could be converted into SG
stars. This is at odds with the ubiquitous presence of a large
fraction of Intermediate SG stars found in all the other clusters that
have been studied with spectroscopic observations (see e.g. Carretta
et al. 2009a).\\
The peculiar star formation history suggested by the results presented
in this paper might be connected to the cluster early dynamical
evolution. Unfortunately, without any information on the orbit of NGC
2419 and the properties of the tidal environment in which this cluster
was embedded during the very early stages of its evolution, it is
difficult to build a model of the early dynamical processes behind the
interruption of the SG  formation event and the mass loss this cluster
must have suffered. \\
At its current galactocentric distance ($R_g\sim 90 kpc$; Harris 1996,
2010 edition), the Jacobi radius of NGC2419 determined by the
strength of the Galactic tidal field at $R_g$ is $r_J\sim 700$
pc. Considering that its estimated King truncation radius is $r_t\sim 220$
pc and its half-mass radius is $r_h\sim 30$ pc, one could infer that
NGC2419 must be the prototype of a cluster evolving in isolation  and
which, therefore, should not have been able to lose any AGB ejecta or
stellar mass. However one can
easily envisage a number of conditions leading to a significantly
different early dynamical history affecting both the gas and the
stellar content. Assuming, for example, the cluster was on an
eccentric orbit with pericenter at 11 kpc, as suggested by
Casetti-Dinescu et al (2009) (see also the discussion in Cohen et
al. 2010), it is possible that an early tidal shock event and/or a
cluster early expansion (triggered by primordial gas and SN ejecta
expulsion as discussed in D'Ercole et al. 2008) in a stronger tidal
field might have affected the SG formation process and caused the loss
of a significant fraction of the initial FG mass.\\
Another possibility is that NGC2419 was located, during the early
stages of its evolution, in the inner regions of a dwarf galaxy and
therefore subject to a stronger tidal field than it is now; also in
this case the early expansion phase as modeled in D'Ercole et
al. (2008) might have led to a strong FG loss. Indeed, Newberg et
al. (2003) suggested that NGC 2419 might have been once part of the
Sagittarius galaxy (see, however, Law \& Majewski, 2010 for a recent
discussion of this possibility).\\
Although only additional studies on the stellar population and on the
orbital properties of NGC2419 will allow to better constrain its
dynamical and star formation history, our results give further support
to previous suggestions (van den berg $\&$ Mackey 2004, Mackey $\&$ van
den bergh 2005, Cohen et al. 2010 and very recently Bruns $\&$ Kroupa 2011) that 
NGC2419 must be the remnant of a much more massive system.   

\section*{Acknowledgments}
We thanks the referee for his useful suggestions that improved the final version of the paper. Financial support for this study was provided by the PRIN MIUR 2007 ``Multiple stellar populations in globular clusters: census, characterization and origin''. EV was supported in part by NASA grant NNX10AD86G.


\begin{thebibliography}{99}

\bibitem[Anderson \& King(2006)]{anderson2006} Anderson, J. \& King, I. 2006,acs rept, 1A
\bibitem[Anderson et al.(2008)]{anderson2008} Anderson, J. et al.,2008, AJ, 135, 2055A
\bibitem[Anderson et al.(2009)]{anderson2009} Anderson, J., Piotto, G., King, I. R., Bedin, L. R., Guhathakurta, P., 2009, ApJ, 697L, 58A
\bibitem[Angulo et al. (1999)]{angulo} Angulo C., Arnould M., Rayet M., et al. 
1999, Nucl. Phys. A, 656, 3 

\bibitem[Bekki \& Norris(2006)]{bekki-norris2006} Bekki, K., \& Norris, 
J.~E.\ 2006, ApJ Letters, 637, L109 

\bibitem[Bekki et al.(2007)]{bekki2007} Bekki, K., Campbell, 
S.~W., Lattanzio, J.~C., \& Norris, J.~E.\ 2007, \mnras, 377, 335 

\bibitem[Bingham et al.(1984)]{bingham1984} Bingham, E.~A., 
Cacciari, C., Dickens, R.~J., \& Pecci, F.~F.\ 1984, \mnras, 209, 765 

\bibitem[\protect\citeauthoryear{Briley et al.}{2002}]{briley2002}
Briley, M., Cohen, J.~G., \& Stetson, P.~B.~2002, ApJ, 579, L17

\bibitem[Briley et al.(2004)]{briley2004} Briley, M.~M., Harbeck, 
D., Smith, G.~H., \& Grebel, E.~K.\ 2004, AJ, 127, 1588  

\bibitem[Brown et al.(2001)]{brown2001} Brown, T.~M., Sweigart, 
A.~V., Lanz, T., Landsman, W.~B., \& Hubeny, I.\ 2001, \apj, 562, 368 

\bibitem[Brown et al.(2010)]{brown2010} Brown, T.~M., Sweigart, 
A.~V., Lanz, T., Smith, E., Landsman, W.~B., \& Hubeny, I.\ 2010, \apj, 718, 1332 

\bibitem[Bruns \& Kroupa (2011)]{kroupa2011} Bruens, R. C. \& Kroupa, P.,  2011, \apj, accepted (arXiv.1101.1306 )

\bibitem[Caloi \& D'Antona(2007)]{caloi2007} Caloi, V., \& D'Antona, F.\ 2007,  A\&A, 463, 949
\bibitem[Caloi \& Mazzitelli(1990)]{caloimazzitelli1990} Caloi, V., 
\& Mazzitelli, I.\ 1990, \aap, 240, 305 
\bibitem[Casetti-Dinescu et al.(2009)]{casetti2009}Casetti-Dinescu, D. I., Girard, T. M., Majewski, S. R., Vivas, A. K., Wilhelm, Ronald, Carlin, Jeffrey L. Beers, Timothy C., van Altena, W. F., 2009, ApJ, 701, 29C
\bibitem[Carretta et al.(2009a)]{carretta2009a} Carretta, E., et al.\ 2009a, A\&A, 505, 117 

\bibitem[Carretta et al.(2009b)]{carretta2009b} Carretta, E., 
Bragaglia, A., Gratton, R., \& Lucatello, S.\ 2009b, A\&A, 505, 139 

\bibitem[Carretta et al.(2009c)]{carretta2009ferro} 
Carretta, E., Bragaglia, A., Gratton, R., D'Orazi, V., \& Lucatello, S.\ 2009c, \aap, 508, 695 

\bibitem[Carretta et al.(2010)]{carretta1851.2010} Carretta, E., et al.\ 
2010, \apjl, 722, L1 

\bibitem[Cassisi et al.(2003)]{cassisi2003} Cassisi, S., Schlattl, 
H., Salaris, M., \& Weiss, A.\ 2003, \apjl, 582, L43 

\bibitem[Catelan et al.(2010)]{catelan2010} Catelan, M., Valcarce, 
A.~A.~R., \& Sweigart, A.~V.\ 2010, IAU Symposium, 266, 281 

\bibitem[Cohen \& Mel{\'e}ndez(2005)]{cohenmelendez2005} Cohen, J.~G., \& 
Mel{\'e}ndez, J.\ 2005, AJ, 129, 303 

\bibitem[Cohen et al.(2010)]{cohen2010} Cohen, J.~G., Kirby, 
E.~N., Simon, J.~D., \& Geha, M.\ 2010, arXiv:1010.0031 

\bibitem[Dalessandro et al.(2008)]{dalessandro2008} Dalessandro, E., 
Lanzoni, B., Ferraro, F.~R., Vespe, F., Bellazzini, M., 
\& Rood, R.~T.\ 2008, \apj, 681, 311 

\bibitem[D'Antona et al.(2002)]{dantona2002} D'Antona, F., Caloi, V., Montalb{\'a}n, J., Ventura, P., \& Gratton, R.\ 2002, \aap, 395, 69 

\bibitem[D'Antona \& Caloi(2008)]{dantonacaloi2008} D'Antona, F., \& Caloi, V.\ 2008, 
MNRAS, 390, 693

\bibitem[D'Antona et al.(2010)]{dantona2010} D'Antona, F., Caloi, 
V., \& Ventura, P.\ 2010, \mnras, 405, 2295 

\bibitem[D'Cruz et al.(1996)]{dcruz1996} D'Cruz, N.~L., Dorman, 
B., Rood, R.~T., \& O'Connell, R.~W.\ 1996, \apj, 466, 359 

\bibitem[D'Cruz et al.(2000)]{dcruz2000} D'Cruz, N.~L., et al.\ 
2000, \apj, 530, 352 


\bibitem[Decressin et al.(2007)]{decressin2007} Decressin, T.,  
Meynet, G., Charbonnel, C., Prantzos, N., \& Ekstr{\"o}m, S.\ 2007, A\&A, 
464, 1029 

\bibitem[D'Ercole et al.(2008)]{dercole2008} D'Ercole, A., 
Vesperini, E., D'Antona, F., McMillan, S.~L.~W., \& Recchi, S.\ 2008, \mnras, 391, 825 

\bibitem[D'Ercole et al.(2010)]{dercole2010} D'Ercole, A., 
D'Antona, F., Ventura, P., Vesperini, E., 
\& McMillan, S.~L.~W.\ 2010, \mnras, 407, 854 

\bibitem[Di Criscienzo et al.(2004)]{dicriscienzo2004} Di Criscienzo, 
M., Marconi, M., \& Caputo, F.\ 2004, \apj, 612, 1092 

\bibitem[Di Criscienzo et 
al.(2010)]{dicriscienzo2010} Di Criscienzo, M., D'Antona, F., \& Ventura, P.\ 2010, \aap, 511, A70 

\bibitem[Di Criscienzo et al.(2011)]{dicriscienzo2011} Di Criscienzo, M., Greco, C., Ripepi, V., Clementini, G., Dall' Ora, M., Marconi, M., Musella, I., Federici, L., Di Fabrizio, 2011, \aj, 141, 81

\bibitem[Dotter  et al.(2007)]{dotter2007} Dotter A., Chaboyer B., Jevremovic D., Baron E., Ferguson J., Sarajedini A., Anderson Y., 2007, AJ, 134, 376
\bibitem[Fenner et al.(2004)]{fenner2004}Fenner,Y., Campbell, S., Karakas, 
A.I., Lattanzio, J.C. \& Gibson, B.K.\ 2004, \mnras, 353, 789

\bibitem[Ferguson et al.(2005)]{ferguson2005} Ferguson  J. W., Alexander  D. R., 
Allard  F.  et al., 2005, ApJ, 623, 585

\bibitem[\protect\citeauthoryear{Formicola et al.}{2004}]{formicola2004}Formicola L., Imbriani G., Costantini H., et al.~2004, Phys. Lett. B, 591, 61
\bibitem[Girardi  et al.(20077)]{girardi2007}Girardi, L., Castelli, F., Bertelli, G., \& Nasi, E. 2007, A\&A, 468, 657
\bibitem[\protect\citeauthoryear{Gratton et al.}{2001}]{gratton2001}
Gratton, R.~G., Bonifacio, P., Bragaglia, A., et al.~2001, A\&A, 369, 87

\bibitem[Han et al.(2003)]{han2003} Han, Z., Podsiadlowski, P., 
Maxted, P.~F.~L., \& Marsh, T.~R.\ 2003, \mnras, 341, 669 


\bibitem[Harris(1996)]{harris1996} Harris, W.E. 1996, AJ, 112, 1487 (updated in 2010 available on  http://physwww.physics.mcmaster.ca/~harris/mwgc.dat)

\bibitem[Iglesias \& Rogers (1996)]{iglesias1996} Iglesias, C.A., Rogers, F.J., 
   ApJ, 464, 943

\bibitem[Ivans et al.(1999)]{ivans1999} Ivans, I.I., Sneden, C., 
   Kraft, R.P., et al., 1999, AJ, 118, 1273

\bibitem[Karakas \& Lattanzio(2007)]{karakas2007} Karakas, A., \& Lattanzio, J.~C.\ 
2007, Publications of the Astronomical Society of Australia, 24, 103 

 \bibitem[Kroupa, P.et al. (1993)]{kroupa1993} Kroupa, Pavel; Tout, Christopher A.; Gilmore, Gerard, 1993, MNRAS, 262, 545
\bibitem[Kroupa, P.(2001)]{kroupa2001} Kroupa, P. 2001, MNRAS, 322, 231K

\bibitem[Law \& Majewski(2010)]{law2010} D.R. Law S.R. Majewski, 2010, ApJ, 718, 1128
\bibitem[Lee et al.(2005)]{lee2005} Lee, Y.-W., et al.\ 2005, \apjl, 621, L57 

\bibitem[Lee et al.(2009)]{lee2009} Lee, J.-W., Lee, J., Kang, 
Y.-W., Lee, Y.-W., Han, S.-I., Joo, S.-J., Rey, S.-C., 
\& Yong, D.\ 2009, \apjl, 695, L78 
\bibitem[\protect\citeauthoryear{{Mackey} \& {van den bergh}}{{Mackey} \& {van den bergh}}{2005}]{mackey05}
{Mackey}, A.~D., \& {van den bergh}, S. 2005, \mnras, 360, 631
\bibitem[Meynet et al.(2006)]{meynet2006} Meynet, G., Ekstr{\"o}m, S., \& Maeder, A.\ 2006, 
A\&A, 447, 623 

\bibitem[Miller Bertolami et al.(2008)]{millerbertolami2008} 
Miller Bertolami, M.~M., Althaus, L.~G., Unglaub, K., \& Weiss, A.\ 2008, \aap, 491, 253 

\bibitem[Milone et al.(2008)]{milone2008} Milone, A.~P., et al.\ 
2008, \apj, 673, 241 

\bibitem[Milone et al.(2010)]{2010ApJ...709.1183M} Milone, A.~P., et al.\ 
2010, \apj, 709, 1183 

\bibitem[Moehler et al.(2002)]{moehler2002} Moehler, S., Sweigart, A.~V., Landsman, W.~B., \& Dreizler, S.\ 2002, \aap, 395, 37 

\bibitem[Moehler et al.(2007)]{moehler2007} Moehler, S., Dreizler, S., Lanz, T., Bono, G., Sweigart, A.~V., Calamida, A., Monelli, M., \& Nonino, M.\ 2007, \aap, 475, L5 


\bibitem[Moni Bidin et al.(2006)]{monibidin2006} Moni Bidin, C., Moehler, S., Piotto, G., Recio-Blanco, A., Momany, Y., \& M{\'e}ndez, R.~A.\ 2006, \aap, 451, 499 

\bibitem[Moni Bidin et al.(2008)]{monibidin2008} Moni Bidin, C., Catelan, M., Villanova, S., Piotto, G., Altmann, M., Momany, Y., \& Moehler, S.\ 2008, Hot Subdwarf Stars and Related Objects, 392, 27 

\bibitem[Newberg et al.(2003)]{newberg2003}Newberg, H. J., Yanny, B., Grebel, E. K., Hennessy, G., Ivezi~, Z., Martinez-Delgado, D.,  2003, ApJ, 596, 191N
\bibitem[\protect\astroncite{Norris}{2004}]{norris2004} 
Norris, J.~E.\ 2004, ApJ Letters, 612, L25 

\bibitem[Pietrinferni et al.(2010)]{pietrinferni2010} Pietrinferni, A., 
Cassisi, S., \& Salaris, M.\ 2010, arXiv:1007.1307 

\bibitem[\protect\astroncite{Piotto et al.}{2005}]{piotto2005} Piotto, G., et al.\ 
2005, ApJ, 621, 777 

\bibitem[Piotto et al.(2007)]{piotto2007} Piotto, G., et al.\ 
2007, ApJL, 661, L53 


\bibitem[Pumo et al.(2008)]{pumo2008} Pumo, M.~L., D'Antona, F., 
\& Ventura, P.\ 2008, \apjl, 672, L25 

\bibitem[Rey et al.(1998)]{rey1998} Rey, S.-C., Lee, Y.-W., 
Byun, Y.-I., \& Chun, M.-S.\ 1998, \aj, 116, 1775 
\bibitem[Renzini(2008)]{renzini2008} Renzini, A. 2008. MNRAS, 391, 354R

\bibitem[Ripepi et al.(2007)]{ripepi07} Ripepi, V., et al.\ 
2007, \apjl, 667, L61

\bibitem[Sabbi et al.(2008)]{sabbi2008} Sabbi, E., et al.\ 2008, \aj, 135, 173 

\bibitem[Sandquist \& Hess(2008)]{sandquist2008} Sandquist, E.~L., \& Hess, J.~M.\ 2008, \aj, 136, 2259 

\bibitem[Shetrone et al.(2001)]{shetrone2001} Shetrone, M.~D., 
C{\^o}t{\'e}, P., \& Sargent, W.~L.~W.\ 2001, \apj, 548, 592 

\bibitem[Siess(2010)]{siess2010} Siess, L.\ 2010, \aap, 512, A10 

\bibitem[Sirianni et al.(2002)]{sirianni2002} Sirianni, M., Nota, 
A., De Marchi, G., Leitherer, C., \& Clampin, M.\ 2002, \apj, 579, 275 

\bibitem[Stolte et al.(2006)]{stolte2006} Stolte, A., Brandner, 
W., Brandl, B., \& Zinnecker, H.\ 2006, \aj, 132, 253 

\bibitem[Trager et al.(1995)]{trager1995}Trager, S. C., King, I. R., \& Djorgovski, S. 1995, \aj, 109, 218 

\bibitem[\protect\citeauthoryear{{van den bergh} \& {Mackey}}{{van den bergh} \& {Mackey}}{2004}]{vdbm04}
{van den bergh}, S.,  \& {Mackey}, A.~D. 2004, \mnras, 354, 713

\bibitem[\protect\citeauthoryear{Ventura et al.}{2001}]{ventura2001} Ventura, P., D'Antona, 
F., Mazzitelli, I., \& Gratton, R.\ 2001, ApJ Letters, 550, L65 


\bibitem[Ventura et al.(2009)]{ventura2009} Ventura, P., Caloi, V., 
D'Antona, F., Ferguson, J., Milone, A., 
\& Piotto, G.~P.\ 2009, \mnras, 399, 934 

\bibitem[Ventura \& D'Antona(2010)]{vd2010} Ventura, P., \& D'Antona, F.\ 2010, arXiv:1009.4527 




\bibitem[Vesperini et al.(2010)]{vesperini2010halo} Vesperini, E., McMillan, S.~L.~W., D'Antona, F., \& D'Ercole, A.\ 2010, \apjl, 718, L112 

\bibitem[Walker(1994)]{walker1994} Walker, A.~R.\ 1994, \aj, 108, 
555 

\bibitem[Whitney et al.(1998)]{whitney1998} Whitney, J.~H., et al.\ 
1998, \apj, 495, 284 

\end{thebibliography}
\end{document}